\documentclass{elsart}

\usepackage{amssymb,natbib,graphicx}

\newcommand{\pder}[2]{\ensuremath{\frac{\partial #1}{\partial #2}}}

\begin{document}

\begin{frontmatter}



\title{Modelling discontinuities and Kelvin-Helmholtz instabilities in SPH}


\author{Daniel J. Price}
\address{School of Physics, University of Exeter, Exeter EX4 4QL, UK}

\ead{dprice@astro.ex.ac.uk}
\ead[url]{http://www.astro.ex.ac.uk/people/dprice/research/kh/}

\begin{abstract}
In this paper we discuss the treatment of discontinuities in Smoothed Particle Hydrodynamics (SPH) simulations. In particular we discuss the difference between integral and differential representations of the fluid equations in an SPH context and how this relates to the formulation of dissipative terms for the capture of shocks and other discontinuities.

This has important implications for many problems, in particular related to recently highlighted problems
in treating Kelvin-Helmholtz instabilities across entropy gradients in SPH.  The specific problems pointed out by Agertz et al. (2007) are shown to be related in particular to the (lack of) treatment of contact discontinuities in standard SPH formulations which can be cured by the simple application of an artificial thermal conductivity term. We propose a new formulation of artificial thermal conductivity in SPH which minimises dissipation away from discontinuities and can therefore be applied quite generally in SPH calculations.

\end{abstract}

\end{frontmatter}
\bibliographystyle{elsart-harv}

\section{Introduction}
 Smoothed Particle Hydrodynamics (SPH) is a Lagrangian particle method for solving the equations of fluid dynamics (for reviews, see \citealt{monaghan05,price04,monaghan92}).
 
  Whilst SPH is widely used in astrophysics, geophysics and engineering applications, recently \citet{agertzetal} have suggested that there are ``fundamental differences'' between SPH and grid-based codes, particularly relating to the simulation of Kelvin-Helmholtz instabilities between two fluids of different densities. Whilst not phrased in terms of ``fundamental differences'', similar problems have been hinted at previously by other authors regarding the treatment of large density gradients in SPH, for example in the context of multi-phase calculations \citep{rt01,mw03}.

  The aim of this paper is to resolve these issues in the broader context of how discontinuities in SPH are treated, starting from an understanding of the difference between integral and differential representations of the fluid equations in SPH (in particular the continuity equation) and thus the need for ``discontinuity-capturing'' terms where differential representations are used.
  
  The paper is structured as follows: Our basic SPH formulation is presented in \S\ref{sec:sphlagrangian} and the treatment of discontinuities is discussed in \S\ref{sec:discont}. We propose a new formulation for artificial thermal conductivity in SPH in \S\ref{sec:discont} which is both effective at resolving contact discontinuities appropriately whilst also minimising the dissipation of thermal energy gradients elsewhere. The general discussion regarding discontinuties is illustrated on shock tube tests presented in \S\ref{sec:sodshock}, on which the effectiveness of the new artificial thermal conductivity formulation is also demonstrated. The problems relating to simulating the Kelvin-Helmholtz instability in SPH are discussed in \S\ref{sec:kh} based on numerical tests. We demonstrate that, whilst there are indeed numerical issues with resolving KH instabilities across contact discontinuities in standard SPH formulations, application of our new term very effectively cures the problem. The results are discussed and summarised in \S\ref{sec:discussion}.

\section{Standard variable$-h$ SPH equations}
\label{sec:sphlagrangian}
 Whilst the standard derivation of the SPH equations from a Lagrangian variational principle has been presented by many authors \citep[e.g.][]{np94,mp01,sh02,monaghan02,pm04b}, it is instructive to repeat the derivation here. We begin with the Lagrangian for a perfect fluid of the form \citep{eckart60,mp01}
\begin{equation}
L = \int \left[ \frac12 \rho v^{2} - \rho u(\rho,s) \right] {\rm dV},
\end{equation}
where $\rho$, ${\bf v}$ and $u$ are the fluid density, velocity and thermal energy per unit mass respectively, the latter assumed to be a function of density and the entropy $s$. This integral is discretised as a sum over SPH particles by replacing the mass element $\rho $dV with the particle mass and the integral by a summation, giving
\begin{equation}
L = \sum_{j=1}^{N} m_{j} \left[\frac12 v_{j}^{2} - u_{j}(\rho_{j},s_{j}) \right].
\label{eq:Lsph}
\end{equation}
The equations of motion for particle $i$ may then be derived using the Euler-Lagrange equations in the form
\begin{equation}
\frac{d}{dt}\left( \pder{L}{{\bf v}_{i}} \right) - \pder{L}{{\bf r}_{i}} = 0,
\label{eq:el}
\end{equation}
provided that all of the quantities in the Lagrangian can be expressed as a function of the particle co-ordinates ${\bf r}$ and their time derivatives $\dot{\bf r} \equiv {\bf v}$.  In this way the exact conservation of momentum, angular momentum and energy is guaranteed in the resultant SPH equations because of the symmetry of the Lagrangian with respect to translations, rotations and time respectively.

The momentum is given straightforwardly, from Equation~(\ref{eq:Lsph}), by
\begin{equation}
\pder{L}{{\bf v}_{i}} = m_{i} {\bf v}_{i}.
\label{eq:dldv}
\end{equation}
The spatial derivatives in the Lagrangian are found by assuming that the entropy is constant and thus that the thermal energy can be expressed as a function solely of the fluid density, giving
\begin{equation}
\pder{L}{{\bf r}_{i}} = -\sum_{j} m_{j} \left.\pder{u_{j}}{\rho_{j}}\right\vert_{s} \pder{\rho_{j}}{{\bf r}_{i}},
\label{eq:dldr}
\end{equation}
where the derivative of thermal energy with respect to density is provided by the first law of thermodynamics at constant entropy, dU$ = -P$dV, where $V = m/\rho$ is the particle volume such that the change in the thermal energy per unit mass is given by
\begin{equation}
{\rm du} = \frac{P}{\rho^{2}} {\rm d\rho}.
\label{eq:dudrho}
\end{equation}

The density in SPH is calculated by sum according to
\begin{equation}
\rho_{i} = \sum_{j} m_{j} W(\vert {\bf r}_{i} - {\bf r}_{j}\vert, h_{i}),
\label{eq:rhosum}
\end{equation}
where in the variable smoothing length formulation $h_{i}$ is in turn assumed to be a function of $\rho_{i}$, in the form
\begin{equation}
h = \eta (m/\rho)^{1/3},
\end{equation}
where $\eta$ specifies the smoothing length in units of the average particle spacing (we use $\eta = 1.2$ throughout this paper). The density summation is thus a non-linear equation for both $\rho$ and $h$ which we solve iteratively using a Newton-Raphson method \citep{pm07}. Taking the time derivative of the density sum, we find
\begin{equation}
\frac{d\rho_{i}}{dt} = \frac{1}{\Omega_{i}}\sum_{j} m_{j} ({\bf v}_{i} - {\bf v}_{j})\cdot \nabla W_{ij} (h_{i}),
\label{eq:rhoevol}
\end{equation}
where $\Omega_{i}$ is a term relating to the derivative of the kernel with respect to the smoothing length. The above is an SPH expression for the continuity equation in the form
\begin{equation}
\frac{d\rho}{dt} = -\rho \nabla\cdot{\bf v}.
\end{equation}

The spatial derivative of density is given by
\begin{equation}
\pder{\rho_{j}}{{\bf r}_{i}} = \frac{1}{\Omega_{j}}\sum_{k} m_{k} \nabla_{i} W_{jk} (h_{j}) (\delta_{ji} - \delta_{ki}),
\end{equation}
giving, via (\ref{eq:dldr}), (\ref{eq:dudrho}), (\ref{eq:dldv}) and (\ref{eq:el}) the SPH equation of motion in the form
\begin{equation}
\frac{d{\bf v}_{i}}{dt} = -\sum_{j} m_{j} \left[ \frac{P_{i}}{\Omega_{i} \rho_{i}^{2}} \nabla W_{ij}(h_{i})  + \frac{P_{j}}{\Omega_{j}\rho_{j}^{2}} \nabla W_{ij}(h_{j})\right].
\end{equation}

 The most common method for integrating the energy equation in SPH is to evolve the thermal energy, which from (\ref{eq:dudrho}) evolves according to
\begin{equation}
\frac{du_{i}}{dt} = \frac{P_{i}}{\Omega_{i} \rho_{i}^{2}} \sum_{j} m_{j} ({\bf v}_{i} - {\bf v}_{j})\cdot \nabla W_{ij}(h_{i}).
\label{eq:dudt}
\end{equation}
Alternatively the total energy $e = \frac12 v^{2} + u$ can be used, which from the Hamiltonian \citep{mp01} has derivative
\begin{equation}
\frac{de_{i}}{dt} = \sum_{j} m_{j} \left[ \frac{P_{i} {\bf v}_{j}}{\Omega_{i} \rho_{i}^{2}}\cdot \nabla W_{ij}(h_{i})  + \frac{P_{j} {\bf v}_{i}}{\Omega_{j}\rho_{j}^{2}}\cdot \nabla W_{ij}(h_{j})\right] .
\end{equation}

A third alternative in the case of an ideal gas is also possible to evolve the entropy function $S = P/\rho^{\gamma}$, which evolves according to
\begin{eqnarray}
\frac{dS}{dt} & = & \frac{\gamma-1}{\rho^{\gamma-1}}\left( \frac{du}{dt} -
\frac{P}{\rho^2}\frac{d\rho}{dt} \right), \nonumber \\
& = & \frac{\gamma-1}{\rho^{\gamma-1}}\left( \frac{du}{dt} \right)_{diss}, \nonumber \\
& = & 0 \hspace{5mm}\textrm{(no dissipation)},
\label{eq:sphentropy}
\end{eqnarray}
where the subscript $(du/dt)_{diss}$ indicates the dissipative part of the evolution of thermal energy. 
The latter has the advantage of placing strict controls on sources of entropy, since
$S$ is purely advected in the absence of dissipative terms \citep{sh02}. It should be noted in the context of our later discussions that the exact advection of entropy is inherently a differential assumption since it relies on the fact that $du - P/\rho^{2} d\rho = 0$ for which there is no corresponding integral conservation law.

 The reader unfamiliar with SPH should also note that in SPH evolving the thermal energy $u$ differs from an evolution involving the conserved total energy $e$ or the entropy $A$ only by the timestepping algorithm. This is quite different to the situation in an Eulerian code where there are additional differences due to advection terms. In SPH an evolution using either $u$, $e$ or $A$ will conserve energy to timestepping accuracy (assuming the terms associated with smoothing length gradients have been properly accounted for as described above).

\section{Discontinuities in SPH}
\label{sec:discont}
 The treatment of flow discontinuities in numerical hydrodynamics has been the subject of a vast body of research over the last 50 years, resulting in the development of a wide range of high accuracy methods for shock capturing schemes mainly applicable in the context of grid-based codes. Related to this has been an understanding of the assumptions necessary for discontinuous solutions to the equations of hydrodynamics (ie. shocks and contact discontinuities) to be captured by the numerical solution. Thus for example, the difference between a finite \emph{volume} scheme and a finite \emph{difference} scheme. Whilst both provide discretisations of the fluid equations on an Eulerian mesh, a finite volume scheme uses a discretisation of the \emph{integral} form of the equations, whilst the starting point for a finite difference scheme is the discretisation of the fluid equations in \emph{differential} form. The problem with the latter is that the assumption that the equations are differentiable immediately excludes the possibility of solutions which have infinite derivatives (requiring additional mechanisms such as dissipative terms in order to capture such solutions), whereas these solutions are not excluded in an integral form.  We discuss below how this relates to formulation of the SPH equations.

\subsection{Density sum versus density evolution}
\label{sec:surf}
 Useful insight into the difference between integral and differential formulations of the fluid equations in SPH may be gained by considering the difference between calculating the density by summation using (\ref{eq:rhosum}) and evolving the density as a fluid variable using (\ref{eq:rhoevol}). It is often assumed that these are two equivalent ways of calculating the density in an SPH calculation since Equation (\ref{eq:rhoevol}) is simply the time derivative of Equation (\ref{eq:rhosum}) and thus that the only distinguishing factor between the two is the cost associated with calculating the density by summation separate to the evaluation of SPH forces \citep[see, e.g.][]{monaghan97}.

 In the light of our discussion above, we expect the two to differ because the density summation represents an integral formulation of the continuity equation (that is, in using the density sum we have nowhere assumed that the density is differentiable) whilst in taking the time derivative we have assumed that the density is a differentiable quantity. The difference between the two can be found by considering the continuity equation written in integral form and smoothed over the local volume using the SPH kernel, ie.
\begin{equation}
\int \left[ \pder{\rho'}{t} + \nabla'\cdot\left(\rho' {\bf v}'\right) \right] W(\vert {\bf r} - {\bf r}' \vert, h) dV' = 0.
\end{equation}
Expanding, we have
\begin{equation}
\pder{}{t} \int \rho' W dV' + \int \nabla' \cdot(\rho' {\bf v}') W dV' = 0,
\end{equation}
where $W = W(\vert {\bf r} - {\bf r}' \vert, h)$. The second term can be expanded further using
\begin{equation}
 \nabla' \cdot\left[\rho' {\bf v}' W \right] = W \nabla'\cdot(\rho' {\bf v}') + \rho' {\bf v}'\cdot \nabla'W,
\end{equation}
giving
\begin{equation}
\pder{}{t} \int \rho' W dV' - \int \rho' {\bf v}' \cdot\nabla' W dV' + \int \nabla' \cdot\left[\rho' {\bf v}' W \right] dV= 0.
\end{equation}
Making use of the antisymmetry of the kernel gradient (ie. $\nabla' W = -\nabla W$) and using Green's theorem to convert the last term from a volume to a surface integral, we have
\begin{equation}
\pder{}{t} \int \rho' W dV' + \int \rho' {\bf v}' \cdot\nabla W dV' + \int \left[\rho' {\bf v}' W \right]\cdot d{\bf S}= 0.
\end{equation}
Replacing the Eulerian time derivative $\partial / \partial t$ with the Lagrangian time derivative, ie. $\partial/ \partial t = d/dt - {\bf v}\cdot\nabla$, we find
\begin{equation}
\frac{d}{dt} \int \rho' W dV' - \int \rho' ({\bf v} - {\bf v'}) \cdot\nabla W dV' + \int \left[\rho' {\bf v}' W \right]\cdot d{\bf S}= 0.
\end{equation}

Finally, we can write the volume integrals as SPH sums by replacing $\rho dV$ with the particle mass and converting the integral to a sum, giving
\begin{equation}
\frac{d}{dt} \sum_{j} m_{j} W_{ij} = \sum_{j} m_{j} ({\bf v}_{i} - {\bf v}_{j})\cdot \nabla W_{ij} - \int \left[\rho' {\bf v}' W \right]\cdot d{\bf S}.
\end{equation}

The above expression clearly shows that a formulation of the SPH continuity equation in integral form would involve \emph{not only} the time derivative of the density sum in the form (\ref{eq:rhoevol}) but also an additional term which appears in the above as a surface integral. This term in general vanishes (that is, the kernel goes to zero at the limits of the integration volume) \emph{except} at boundaries, or equivalently, flow discontinuities. Thus we expect that SPH formulations which evolve the continuity equation in the form (\ref{eq:rhoevol}) will differ from formulations which utilise the density sum at such discontinuities, the latter of which should require no special treatment. That this is indeed the case is demonstrated in numerical tests presented in \S\ref{sec:tests}. In the context of our variable smoothing length SPH formulation, it is important to note that a true ``integral form'' of the density sum is \emph{only} obtained when the smoothing length is \emph{also} obtained directly from the summation via iteration of the smoothing length-density relation discussed in \S\ref{sec:sphlagrangian} (as opposed to simply evolving $h$ separately using a differential form of the continuity equation).

 Regarding discontinuities in other variables (apart from density), the appearance of surface integrals also provides some insight into where discontinuities, e.g. in velocity ``go missing'' when deriving the SPH equations from a Lagrangian presented in \S\ref{sec:sphlagrangian}. For example, in using the Euler-Lagrange equations (\ref{eq:el}) we have implicitly assumed that the variation in the action vanishes at the surface of the integration volume (that is, certain surface integrals involving the Lagrangian are assumed to vanish). An assumption of differentiability is also apparent from the fact that the Euler-Lagrange equations contain derivatives with respect to particle coordinates and velocity and can only be derived in this form by assuming that the variation in the action vanishes at the surface of the integration volume. The reader will thus note that in our SPH derivation in \S\ref{sec:sphlagrangian} a differentiated (and thus assumed differentiable) version of the density sum was used in finding the equations of motion, leading to the inevitable consequence that, whilst the continuity equation can be solved in an integral form using the density summation, the SPH momentum and energy equations derived above are clearly differential.

\subsection{Artificial dissipation terms}
  The discussion above leads to an obvious corollary, namely given that discontinuities have ``gone missing'' from the SPH by the assumption of differentiability, how should they be recovered in the numerical solution? The simplest approach is to add dissipation terms to the SPH equations which diffuse discontinuities on the smoothing scale such that they are resolved by the numerical method (and thus no longer ``discontinuous'').  A general formulation of such dissipative terms was presented by \citet{monaghan97} in a comparison of SPH to grid-based codes incorporating Riemann solvers. Whilst the usual approach taken in SPH is to simply add an artificial viscosity term to the momentum equation, \citet{monaghan97} noted that, by analogy with Riemann solvers, the evolution equation for \emph{every} conservative variable should contain a corresponding dissipation term in it's evolution related to jumps in that variable, leading naturally to formulations of dissipative terms for ultra-relativistic shocks \citep{cm97} and for Magnetohydrodynamics (MHD) \citep{pm04a,pm05} in SPH. 
  

\subsubsection{Hydrodynamics}
  For a non-relativistic gas the dissipation terms for the evolved variables in conservative form (namely the conserved momentum and energy per unit mass,  ${\bf v}$ and $e = \frac12 v^{2} + u$ respectively) take the form \citep{monaghan97}
\begin{eqnarray}
\left(\frac{d{\bf v}_{i}}{dt}\right)_{diss} & = & \sum_j m_j \frac { \alpha v_{sig} ({\bf v}_i -
{\bf v}_j ) \cdot \hat{\bf r}_{ij}}{\bar{\rho}_{ij} } \overline{\nabla_i W_{ij}}, \label{eq:av} \\
\left(\frac{de_i}{dt}\right)_{diss} & = & \sum_j m_j \frac{(e^*_i - e^*_j)}{\bar{\rho}_{ij} } \hat{\bf r}_{ij} \cdot \overline{\nabla_i W_{ij}},
\label{eq:ave}
\end{eqnarray}
where the bar over the kernel refers to the fact that the kernel must be symmetrised with respect to $h$, ie.
\begin{equation}
\overline{\nabla W_{ij}} = \frac12 \left[\nabla W_{ij}(h_{i}) + \nabla W_{ij}(h_{j}) \right],
\end{equation}
and the energy variable $e_{i}^{*} =  \frac12 \alpha v_{sig}({\bf v}_{i} \cdot \hat{\bf r}_{ij})^{2} + \alpha_{u}  v_{sig}^{u} u_{i} $ refers to an energy including only components along the line of sight joining the particles with different parameters ($\alpha$, $\alpha_{u}$) specifying the dissipation applied to each component. The choice of signal speed $v_{sig}$ is discussed below (\S\ref{sec:vsig}). Note, however, that in this paper we have deliberately distinguished between the signal velocities used for the kinetic energy term $v_{sig}$ and that used for the thermal energy term ($v_{sig}^{u}$), for reasons that will become clear. This differs from previous formulations (e.g. \citealt{monaghan97,pm04a,pm05}) which have assumed that the same signal velocity is used to treat jumps in all variables.

 Equation (\ref{eq:av}) in the \citet{monaghan97} formulation provides an artificial viscosity term similar to earlier SPH formulations (e.g. \citealt{monaghan92} -- the two formulations differ only by a factor of $h/\vert r_{ij} \vert$). Equation (\ref{eq:ave}) is more interesting, since (as discussed by \citealt{monaghan97}) it shows that the evolution of the total energy should contain not only a term relating to jumps in kinetic energy (ie. the thermal energy contribution from the viscosity term) but also a term relating to jumps in thermal energy. This is more explicitly obvious if we consider the evolution of the thermal energy resulting from the above formulation, ie.
\begin{equation}
\frac{du}{dt} = \frac{de}{dt} - {\bf v}\cdot \frac{d{\bf v}}{dt},
\end{equation}
which, using (\ref{eq:av}) and (\ref{eq:ave}) gives
\begin{eqnarray}
\left(\frac{du}{dt}\right)_{diss} & = & -\sum_j \frac{m_{j}}{\bar{\rho}_{ij}}\left\{ \frac{1}{2}\alpha v_{sig}
({\bf v}_{ij}\cdot\hat{\bf r}_{ij})^2 \right. \nonumber \\
& & \left.\phantom{\frac{1}{2}} +  \alpha_u v_{sig}^{u} (u_i - u_j) \right\} \hat{\bf r}_{ij}\cdot \nabla_i W_{ij}. \label{eq:udiss}
\end{eqnarray}
 The term involving $(u_{i} - u_{j})$ provides an artificial thermal conductivity which acts to smooth discontinuities in the thermal energy. \emph{The need for such an artificial thermal conductivity contribution in order to resolve discontinuities in thermal energy is almost universally ignored in SPH formulations.}
 
  The effect of applying different types of dissipation to specific discontinuities is discussed in the MHD case by \citet{pm05} and in the hydrodynamic case by \citet{price04}. The point made in these papers is that every physical discontinuity requires an appropriate treatment. For example in hydrodynamics, shocks are treated by the application of artificial viscosity terms but accurate treatment of contact discontinuities requires the addition of artificial thermal conductivity to treat the jump in thermal energy. In the MHD case discontinuities in the magnetic field are treated separately by the application of artificial resistivity.  We discuss the hydrodynamic case in more detail below and in the shock tube tests presented in \S\ref{sec:sodshock}. In \S\ref{sec:kh} we show how these results have a direct bearing on the problems encountered when trying to simulate Kelvin-Helmholtz instabilities across density jumps in SPH.
 
\subsubsection{Interpretation of dissipative terms}
\label{sec:monaghan97}
 The dissipation terms introduced by \citet{monaghan97} can be interpreted more generally as ``discontinuity capturing'' terms. Interpreted as such, for any \emph{conservative} variable (ie. such that $\sum_{j} m_{j} dA_{j}/dt = 0$) that is evolved via a differential equation one would expect to add a dissipation term of the general form (for a scalar quantity $A$)
\begin{equation}
\left(\frac{dA_{i}}{dt}\right)_{diss} = \sum_{j} m_{j} \frac{\alpha_{A} v_{sig}}{\bar{\rho}_{ij}} (A_{i} - A_{j}) \hat{\bf r}_{ij}\cdot\nabla W_{ij}
\label{eq:gendiss}
\end{equation}
where $\alpha_{A}$ is a parameter of order unity specifying the amount of diffusion to be added to the evolution of $A$. The interpretation of (\ref{eq:gendiss}) can be seen by considering the SPH expression for the Laplacian in the form \citep[e.g.][]{brookshaw85}
\begin{equation}
(\nabla^{2} A)_{i} = 2  \sum_j m_j \frac{(A_i - A_j)}{\rho_j } \frac{F_{ij}}{\vert r_{ij} \vert}
\label{eq:sphlaplacian}
\end{equation}
where the scalar function $F_{ij}$ is the dimensionless part of the kernel gradient such that $\nabla W_{ij} = \hat{\bf r}_{ij} F_{ij}$ and thus $\hat{\bf r}_{ij}\cdot\nabla W_{ij} \equiv F_{ij}$. We then see that (\ref{eq:gendiss}) is simply an SPH representation of a diffusion term of the form
\begin{equation}
\left(\frac{dA}{dt}\right)_{diss} \approx \eta \nabla^{2} A,
\end{equation}
with a diffusion parameter $\eta$ proportional to the resolution length\footnote{Note that whilst the resolution length appears as the particle spacing, this is similar to the smoothing length since within the kernel radius $\vert r_{ij} \vert /h < 2$}
\begin{equation}
\eta \propto \alpha v_{sig} \vert r_{ij} \vert.
\end{equation}

\subsubsection{Choosing the signal velocity}
\label{sec:vsig}
 In previous formulations \citep{monaghan97,pm04b,pm05} the signal speed $v_{sig}$ used in both the artificial viscosity and conductivity terms is chosen to be an estimate of the magnitude of the maximum signal velocity between a particle pair, an estimate for which (for non-relativistic hydrodynamics) is given by \citep{monaghan97}
\begin{equation}
v_{sig} = \frac12 \left[ c_i + c_j - \beta  {\bf v}_{ij} \cdot \hat{\bf r}_{ij} \right],
\label{eq:vsig}
\end{equation}
where $c$ is the sound speed and generally $\beta = 2$. However, whilst using a signal velocity based on the sound speed and relative particle velocities is appropriate at shocks (which travel at the sound speed and involve strong compression), it is not clear that the same signal velocity should be used to treat contact discontinuities (where there is no compression and the motion is at the post-shock velocity). A good example is to consider the simplest case of two regions with different densities and temperatures in pressure equilibrium. Applying artificial thermal conductivity using a signal velocity proportional to the sound speed would result in a steady diffusion of the initial discontinuity in thermal energy, which as $t\to \infty$ would have completely eliminated the temperature gradient between the two regions.

 A much better approach suggested by the shock tube results discussed in \S\ref{sec:sodshock} is to apply artificial conductivity \emph{only} in order to eliminate spurious pressure gradients across contact discontinuities. In order to do so we require a signal velocity which vanishes when the pressure difference between a particle pair is zero. We propose the following
\begin{equation}
v^{u}_{sig} = \sqrt{\frac{\vert P_{i} - P_{j} \vert}{\bar\rho_{ij}}},
\label{eq:vsigu}
\end{equation}
which is constructed to have dimensions of velocity and to be zero once pressure equilibrium is reached. We find that this is a very effective approach to introducing artificial thermal conductivity into SPH in a controlled manner to appropriately treat contact discontinuities without the side-effect of unwanted diffusion elsewhere. This is particularly the case in the Kelvin-Helmholtz tests discussed in \S\ref{sec:kh}.

 
\subsubsection{Reducing dissipation away from discontinuities}
\label{sec:switches}
 The key problem with using dissipative terms for capturing discontinuities is that such terms also tend to dissipate gradients which are not purely discontinuous. This is a particular problem in relation to artificial thermal conductivity, since whilst shocks are continually steepened by the propagating wave, a gradient in thermal energy, once diffused, will remain diffused forever. The art is therefore to come up with well-designed switches that turn the dissipation terms off away from discontinuities.
 
 In this paper we adopt the artificial viscosity switch suggested by \citet{mm97}, where the viscosity parameter $\alpha$ is different for every particle and evolved according to a simple source and decay equation of the form
\begin{equation}
\frac{d\alpha_{i}}{dt} = -\frac{\alpha_{i}-\alpha_{min}}{\tau_{i}} + \mathcal{S}_{i},
\label{eq:dalphadt}
\end{equation}
such that in the absence of sources $\mathcal{S}$, $\alpha$ decays to a value
$\alpha_{min}$ over a timescale $\tau$. The timescale $\tau$ is calculated
according to
\begin{equation}
\tau_{i} = \frac{h_{i}}{\mathcal{C}v_{sig}},
\end{equation}
where $h$ is the particle's smoothing length, $v_{sig}$ is the maximum signal
propagation speed for particle $i$ (ie. the maximum over pairs involving $i$ of the pairwise $v_{sig}$ defined in Equation~\ref{eq:vsig}) and $\mathcal{C}$ is a dimensionless parameter which we set to $\mathcal{C}=0.1$ which means that the value of $\alpha$ decays to $\alpha_{min}$ over $\sim 5$ smoothing lengths. In general we also impose a maximum value of $\alpha_{max} = 1$ throughout the evolution. In the dissipation terms (\ref{eq:av}) and (\ref{eq:udiss}) the average value on the particle pair $\alpha = 0.5(\alpha_{i} + \alpha_{j})$ is used to maintain symmetry.

The source term $\mathcal{S}$ is chosen such that the artificial dissipation grows
as the particle approaches a shock front. We use, as in \citet{mm97},
\begin{equation}
\mathcal{S} = \mathrm{max}(-\nabla\cdot{\bf v}, 0),
\label{eq:alphasource}
\end{equation} 
such that the dissipation grows in regions of strong compression. 

A similar switch for the artificial thermal conductivity was introduced by \citet{pm05} (see also \citealt{price04}), where the controlling parameter  $\alpha_{u}$ is evolved according to (\ref{eq:dalphadt}) with the minimum value $\alpha_{u,min}$ set to zero and a source term based on a second derivative of the thermal energy,
\begin{equation}
\mathcal{S_{i}} = \frac{h_{i} \vert \nabla^2 u \vert_{i} }{\sqrt{u_{i} + \epsilon}},
\label{eq:source1}
\end{equation} 
where $h$ is the smoothing length, $\epsilon$ is a small number to prevent divergences for small $u$ and the second derivative term is computed using the standard SPH formulation for the Laplacian (Equation \ref{eq:sphlaplacian}). Note that the decay timescale $\tau$ in this case is kept the same as for the viscosity, ie. using (\ref{eq:vsig}). In this paper we find that the combination of our new $v_{sig}^{u}$ and the above switch are very effective at turning the conductivity off away from discontinuities (in fact with the new $v_{sig}^{u}$ the switch is almost unnecessary).


\subsection{Alternative approaches}
\label{sec:rt01}
  \citet{rt01} (hereafter RT01) suggested an alternative approach to dealing with problems with contact discontinuities in multiphase calculations based on a smoothed estimate of pressure. In their formulation the SPH force equation takes the form (modified here slightly to assume an adiabatic equation of state and in the form of the symmetry of the kernels with respect to the smoothing length)
\begin{equation}
\frac{d{\bf v}_{i}}{dt} = (1 - \gamma) \sum_{j} m_{j} \left[ \frac{u_{j}}{\langle \rho_{i} \rangle} \nabla W_{ij}(h_{i})  + \frac{u_{i}}{\langle \rho_{j}\rangle} \nabla W_{ij}(h_{j})\right],
\end{equation}
which, translated, is an SPH form of
\begin{equation}
\frac{d{\bf v}_{i}}{dt} = -\left[ \frac{\nabla P}{\rho} + \frac{P}{\rho} \nabla 1 \right].
\end{equation}
The mean density $\langle \rho \rangle$ used in the force term is derived from a smoothed pressure estimate, in the form
\begin{equation}
\langle \rho_{i} \rangle = \frac{\langle P_{i} \rangle}{(\gamma - 1) u_{i}} = \frac{\sum_{j} m_{j} u_{j} W_{ij} (h_{i})}{u_{i}}.
\end{equation}
 The motivation behind this formulation was to compute the pressure force without the SPH density explicitly appearing in the equations, in order to better handle pressure profiles across strong density gradients.  We compare the results of this formulation with the standard SPH equations (and to our formulation using artificial thermal conductivity) in \S\ref{sec:kh} (note that for the comparison in this paper we use the smoothing length calculated by iteration with the usual SPH density estimate as described in \S\ref{sec:sphlagrangian}). We indeed find an improvement in pressure continuity at discontinuities using their formulation, though at the expense of considerable particle noise at the interface.
 
 \citet{mw03} have also proposed a modified SPH formulation for multiphase flows, in the form of several somewhat \emph{ad-hoc} criteria for excluding particles from one another's neighbour lists. It is however difficult to see how their method can be adopted into a consistent Lagrangian formulation of the SPH equations, particularly since the total energy and/or momentum could arbitrarily change between timesteps as particle pairs are excluded (or not) from the calculation. Furthermore the first of their exclusion criteria is that the density contrast should be greater than 10, whereas problems with resolving the Kelvin-Helmholtz ability occur for much smaller density ratios (as demonstrated in \S\ref{sec:kh}) where their proposed modifications would have no effect. Thus we do not consider their formulation any further here.

\section{Tests}
\label{sec:tests}
Whilst there are many problems for which the incorporation of an artificial thermal conductivity in SPH is a crucial requirement (e.g. \citet{rp07} find significantly improved results on Sedov blast wave tests when such a term is incorporated), conciseness limits us to the consideration of two particular cases. We first consider a one dimensional shock tube problem in order to illustrate the ideas presented in this paper, before applying the method to simulations of the Kelvin-Helmholtz instability across a contact discontinuity (similar to those performed by \citet{agertzetal}), to show that the problem highlighted by these authors is very effectively cured by our new artificial thermal conductivity formulation. 

\subsection{One dimensional shock tube problem}
\label{sec:sodshock}
 For the first test we consider a one dimensional shock tube problem where an initial discontinuity in pressure and density is set up at the origin. We set up the problem with conditions to the left of the discontinuity given by $(\rho_{L}, P_{L}, v_{L}) = (1.0, 1.0, 0.0)$ and conditions to the right given by $(\rho_{R}, P_{R}, v_{R}) = (0.125, 0.1, 0.0)$ with $\gamma = 5/3$. Equal mass particles are used such that the particle separation varies according to the density. We choose the spacing to the right of the origin  as $\Delta_{R} = 0.01$ which results in a spacing to the left of the origin of $\Delta_{L} = 0.00125$, giving a total of $450$ particles in the domain $x = [-0.5, 0.5]$. Most importantly we set up the problem using \emph{unsmoothed} initial conditions in \emph{all} variables. Using smoothed initial conditions effectively smoothes the initial contact discontinuity throughout the evolution and the problems discussed below do not become apparent (for a discussion of this point, see \citealt{price04}). 
 
  It may be noticed in passing that the need for artificial thermal conductivity in this problem has already also been discussed in some detail by \citet{price04}, which we are essentially repeating here because it has a direct bearing on the behaviour of Kelvin-Helmholtz instabilities across contact discontinuities and thus to the \citet{agertzetal} ``problem''.
  
\begin{figure}
\includegraphics[angle=270,width=\columnwidth]{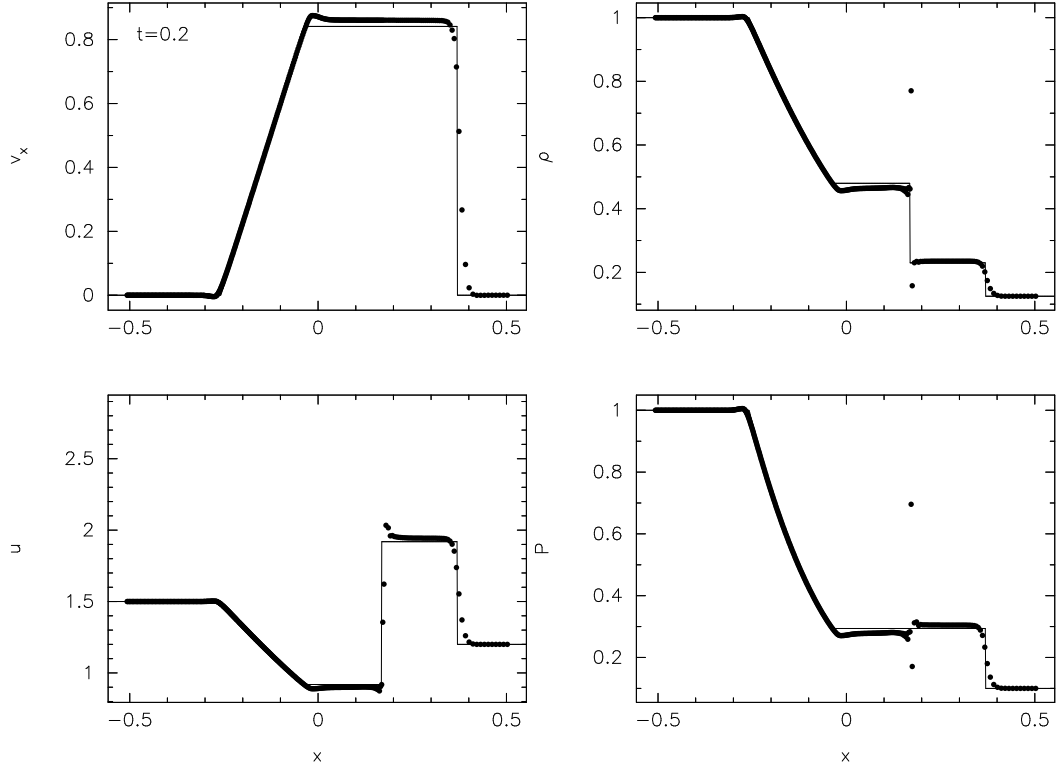}
\caption{Results of the one dimensional shock tube problem with purely discontinuous initial conditions using differential forms of the SPH equations. Artificial viscosity has been applied uniformly and the density (and smoothing length) have been evolved using the SPH form of the continuity equation. The contact discontinuity is unresolved in the density and results in a spurious `blip' in the pressure profile and a slight overshoot in the thermal energy. The shock is smoothed to several resolution lengths by the artificial viscosity term.}
\label{fig:sodcty}
\end{figure}

  To proceed in the order of the discussion given in \S\ref{sec:discont}, we will first begin by examining the difference between calculating the density in SPH by direct summation in the form (\ref{eq:rhosum}) and by evolving the continuity equation in the form (\ref{eq:rhoevol}). We first consider the evolution in ``differential form'', that is, using the continuity equation. We apply only artificial viscosity in the form (\ref{eq:av}) with a constant coefficient $\alpha = 1$ (to keep things simple). The results are shown at $t=0.2$ in Figure~\ref{fig:sodcty}, where the SPH particles are indicated by the points and may be compared to the exact solution given by the solid line. Whilst the shock is smoothed to several resolution lengths by the artificial viscosity term, immediately visible is a large `blip' in the pressure profile around the contact discontinuity because of the unresolved gradient in density and a slight overshoot in the thermal energy at the same location.

\begin{figure}
\includegraphics[angle=270,width=\columnwidth]{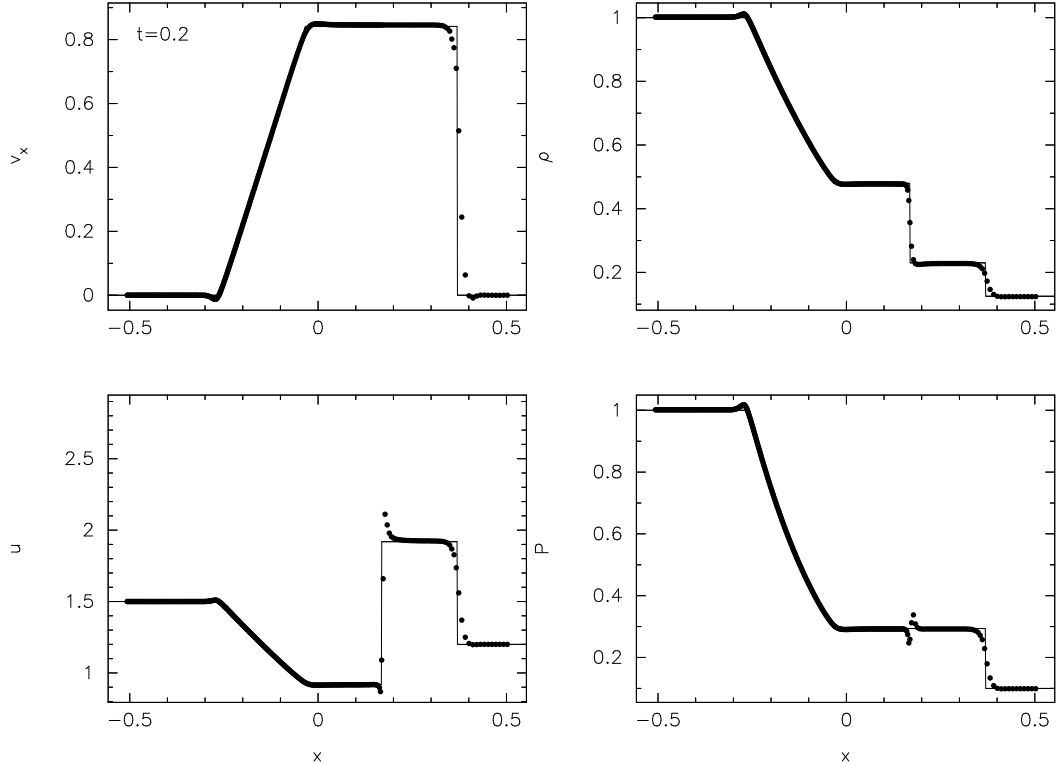}
\caption{As in Figure~\ref{fig:sodcty} but where the density has been calculated by direct summation which represents an integral formulation of the continuity equation in SPH. The density gradient at the contact discontinuity is much closer to the exact solution, although there remain problems with the thermal energy gradient and hence the pressure.}
\label{fig:rhosum}
\end{figure}

 A significant improvement in the density gradient is obtained by using the density summation instead of evolving the continuity equation, shown in Figure~\ref{fig:rhosum}. In the light of the discussion presented in \S\ref{sec:surf} we expect no problem with the density discontinuities in this case because the density summation represents an integral formulation of the continuity equation (put another way, the density summation is the exact solution to the SPH continuity equation which finds the change in density due to a change in the particle positions). That indeed this is the case is demonstrated in Figure~\ref{fig:rhosum}, which shows that using the density summation the density gradient is well resolved across the contact discontinuity. However, a spurious blip in the pressure is still apparent due to the overshoot in thermal energy.

 The reader who is familiar with so-called ``high-resolution shock capturing'' grid based codes may object at this point that the shock profiles in Figure~\ref{fig:rhosum} appear excessively smoothed compared to the best grid-based results. Although it may be countered that the shock width in a numerical simulation is a somewhat meaningless quantity (since the real shock width is many orders of magnitude smaller than the resolution length of any numerical code), recent efforts \citep{inutsuka02,cw03} have shown that such methods can also be utilised in an SPH context. The main objection to doing so (raised for example by \citealt{monaghan97}) is that Godunov schemes are more difficult to implement when the equation of state is non-trivial (though methods exist -- e.g. \citealt{cg85}), whereas artificial dissipation terms are easily applied in all contexts regardless of the degree of complication in the physics or the equation of state [for example in an SPH context, the M97 formulation of the artificial dissipation terms presented above has been readily applied to magnetohydrodynamics \citep{pm04a,pm05} and to ultra-relativistic flows \citep{cm97}].

\begin{figure}
\includegraphics[angle=270,width=\columnwidth]{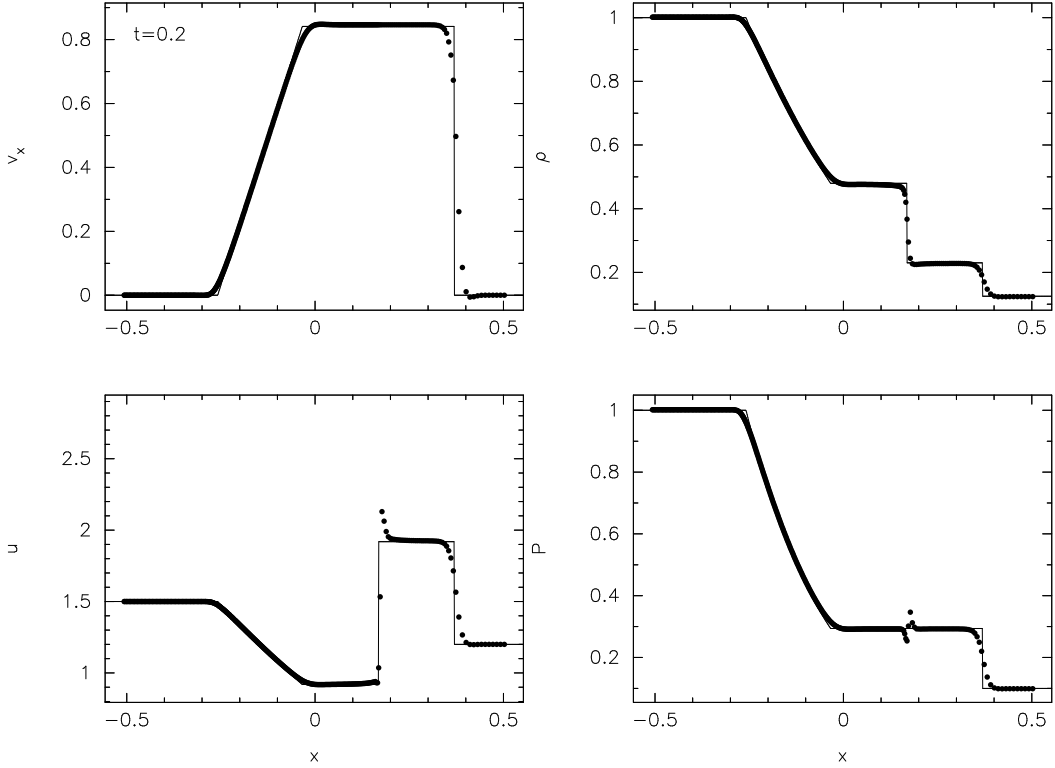}
\caption{As in Figure~\ref{fig:rhosum} (with the density calculated by direct summation) but using a Godunov-SPH scheme instead of applying artificial viscosity. The results are almost indistinguishable from Figure~\ref{fig:rhosum} except for a slightly larger smearing of the rarefaction in the present case (where artificial viscosity is not normally applied). The same problem regarding the thermal energy gradient at the contact discontinuity is apparent even in this case.}
\label{fig:sodgsph}
\end{figure}

 To show that our key conclusions are not affected by the particular form of the ``discontinuity capturing terms'', Figure~\ref{fig:sodgsph} shows the results using the simplest (first-order) Godunov-SPH scheme suggested by \citet{cw03} (that is, where we have simply replaced $P_{i}$ and $P_{j}$ in the SPH force equation with $P^{*}$, the solution to the Riemann problem treating particles $i$ and $j$ as the left and right states, also using $P^{*}$ in the thermal energy equation) instead of artificial viscosity. The results are almost indistinguishable to those using artificial viscosity (Figure~\ref{fig:rhosum}) apart from a slightly larger smearing of the rarefaction (no artificial viscosity is applied to rarefactions in standard SPH) but critically also produce a similarly discontinuous pressure at the contact discontinuity (a similar result may be observed in the tests performed by \citealt{cw03}). This is easily understood since in the simplest Godunov-SPH formulation only the jump in pressure has been addressed by the scheme and pressure should be constant across a contact discontinuity. It is therefore apparent that even in this case some additional treatment is required to address discontinuities in thermal energy. \citet{inutsuka02} finds similar results using Godunov-SPH unless an integral formulation for the momentum equation is used based on a polynomial interpolation of density.

\begin{figure}
\includegraphics[angle=270,width=\columnwidth]{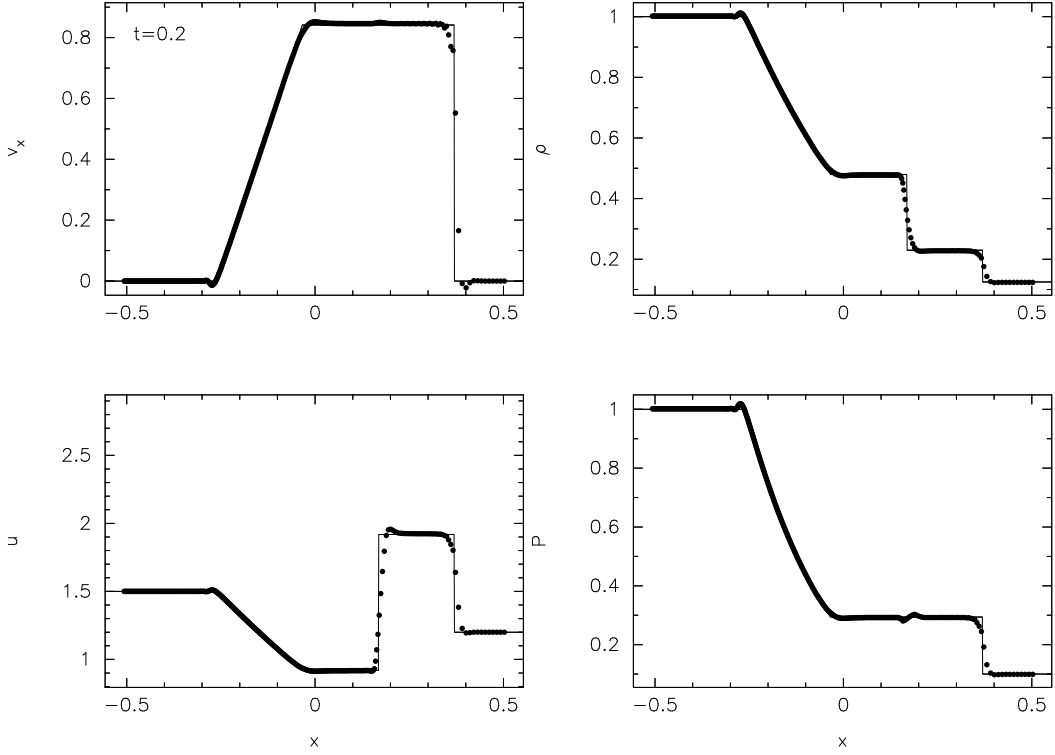}
\caption{As in Figure~\ref{fig:rhosum} (with the density calculated by direct summation) but also with an artificial thermal conductivity term added using the new pressure-jump dependent signal velocity. Artificial viscosity has been applied using the switch discussed in \S\ref{sec:switches}. The density and thermal energy gradients are in this case both resolved across the contact discontinuity and most importantly the pressure across the contact is continuous, unlike in any of our previous results.}
\label{fig:sodfinal}
\end{figure}

 Thus, regardless of the exact form of the ``discontinuity capturing terms'', the key point remains that \emph{every} evolution equation in SPH resulting from a differential formulation requires a term which treats discontinuities in that variable. In the present case the missing piece is in the energy equation which takes the form of an artificial thermal conductivity. The results of this problem using the density summation and including an artificial thermal conductivity term of the form (\ref{eq:udiss}) using the new signal velocity (which we expect to diffuse the gradient in thermal energy until the pressure is continuous) are shown in Figure~\ref{fig:sodfinal}. The artificial viscosity has also been applied using the switch given in \S\ref{sec:switches} such that $\alpha$ is a time variable parameter which responds to the convergent velocities, demonstrating that the use of such a switch nonetheless supplies the necessary dissipation of kinetic energy at a shock front. From Figure~\ref{fig:sodfinal} we see that in this case all of the gradients in all of the variables follow the exact solution and, most importantly, unlike in all of our previous results, the pressure is now continuous (and constant) across the contact discontinuity. The continuity of pressure turns out to be of crucial importance in simulating Kelvin-Helmholtz instabilities across such discontinuities (see below).



\subsection{Kelvin-Helmholtz instabilities across a contact discontinuity}
\label{sec:kh}
 Finally, we consider the problem of Kelvin-Helmholtz instabilities across a density jump, on which recent problems have been highlighted by \cite{agertzetal}. We set up the problem in two spatial dimensions, using equal mass particles in a periodic box in the domain $-0.5 < x < 0.5$, $-0.5 < y < 0.5$. The particles are placed on a uniform cubic lattice in separate regions such that $\rho = 2$ for $\vert y \vert < 0.25$ and $\rho = 1$ elsewhere, where to ensure symmetry in the initial conditions we first set up particles in the domain $y > 0$ and then reflect the initial particle distribution across the $y=0$ axis. The regions are placed in pressure equilibrium with $P= 2.5$ and we use an ideal gas equation of state $P = (\gamma -1 )\rho u$ with $\gamma = 5/3$ such that there is a corresponding jump in thermal energy at the contact. As in the one dimensional tests, the initial gradients in density and thermal energy are not smoothed in any way (although we do first calculate the density by summation before setting the thermal energy to ensure that pressure is at least continuous in the initial conditions). Periodic boundary conditions are implemented using ghost particles.
  
 A cartesian shear flow is setup in the $x-$direction with velocity $v_{x} = 0.5$ for $\vert y \vert < 0.25$ and $v_{x} = -0.5$ elsewhere. Such a configuration is known to be unstable to the Kelvin-Helmholtz instability at all wavelengths. In this paper we seed the instability at a particular wavelength by applying a small velocity perturbation in the $y$ direction given by
\begin{equation}
v_{y} = \left\{ \begin{array}{ll}
 A \sin{[-2\pi (x+0.5)/\lambda]} & \vert y - 0.25 \vert < 0.025,  \\
 A \sin{[2\pi (x+0.5)/\lambda]} & \vert y + 0.25 \vert < 0.025.
\end{array} \right.
\end{equation}
where we use $\lambda = 1/6$ and $A = 0.025$.

 A textbook analysis of the incompressible Kelvin-Helmholtz instability (e.g. \citealt{choudhuri98}), applicable here since the velocities are smaller than the sound speed, shows that the characteristic growth timescale of the instability between two shearing layers of densities $\rho$ and $\rho'$ is given by
\begin{equation}
\tau_{KH} = 2\pi/\omega,
\end{equation}
where
\begin{equation}
\omega = \frac{2\pi}{\lambda}\frac{(\rho \rho')^{1/2} \vert v_{x} - v_{x}' \vert}{(\rho + \rho')}.
\end{equation}
For the setup described above, $\rho =1, \rho' = 2, v_{x} = 0.5$ and $v_{x} = -0.5$, giving $\tau_{KH} = 0.35$ in the units of our simulation for a perturbation of wavelength $\lambda = 1/6$. For the case where the density ratio is 10:1 (ie. $\rho = 1, \rho' = 10$) we have $\tau_{KH} = 0.58$ for $\lambda = 1/6$.

 We define the resolution in the simulations according to the particle spacing in the least-dense region. For the 2:1 density ratio we show results using particle spacing of $\Delta = 1/256$ and $\Delta = 1/512$ in this region, resulting in a total of $97,928$ and $393,160$ particles respectively. For the 10:1 case we show results using $\Delta = 1/256$ in the less-dense region, resulting in a total of $359,604$ particles. We find that the resolution in the low-density region has an important effect on the resolution of the vortex rolls created by the Kelvin-Helmholtz instability. Whilst it would be possible to set up the problem using unequal mass particles (and thus equal resolution in both regions), it would be rather contrived to do so since a density gradient involving equal mass particles is the situation which naturally occurs in SPH simulations.
 
\begin{figure*}
\includegraphics[width=\textwidth]{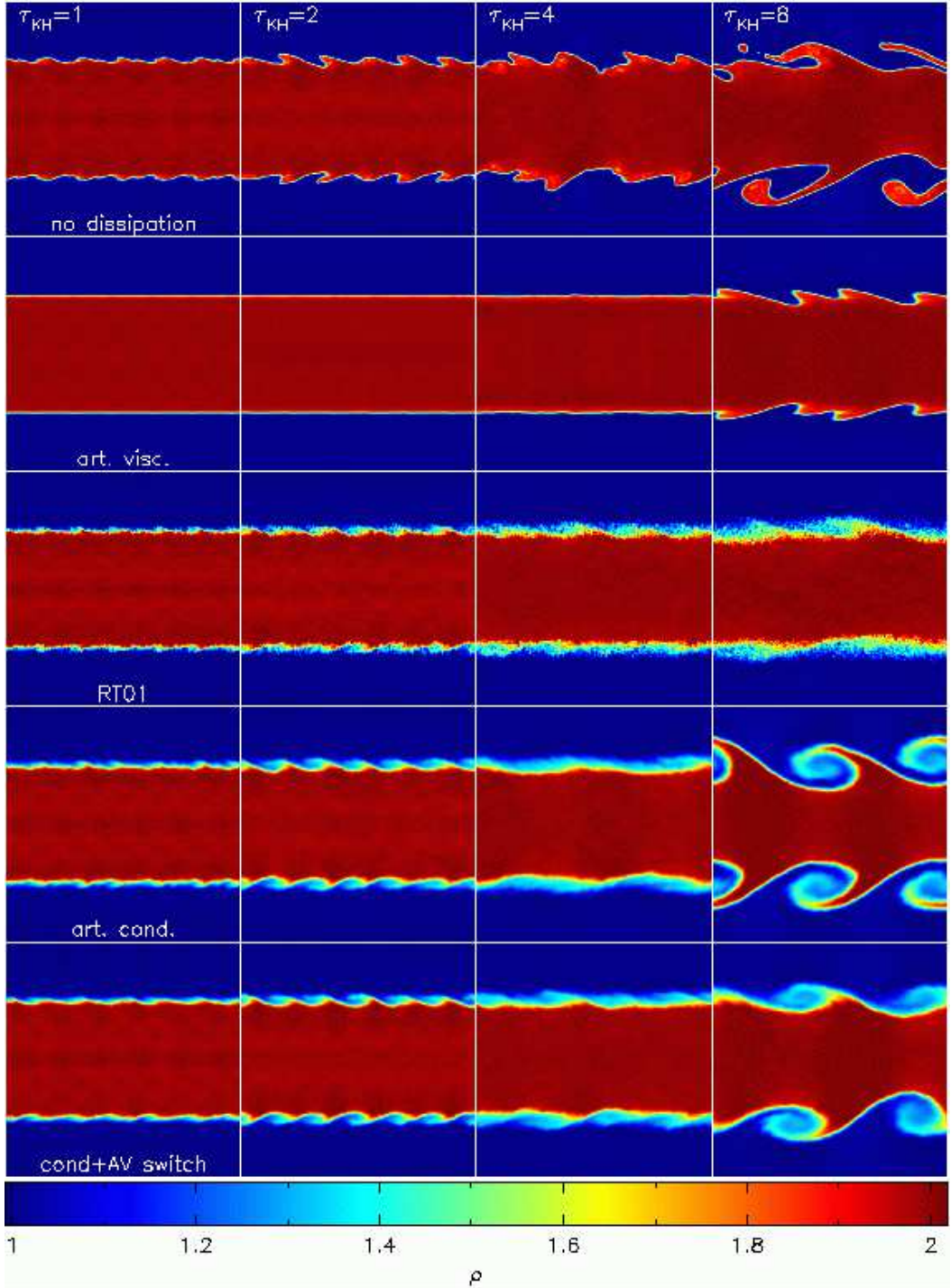}
\caption{Results of the Kelvin-Helmholtz instability test using a density ratio of 2:1 and an initial particle spacing of $\Delta = 1/256$ in the least-dense component. Results are shown using (from top to bottom): 1) no artificial viscosity or thermal conductivity, 2) artificial viscosity applied uniformly, 3) using only artificial thermal conductivity applied using our new signal velocity and 4) with both artificial viscosity and thermal conductivity, where viscosity is applied exactly as in the shock tube problem presented in Figure~\ref{fig:sodfinal}.}
\label{fig:kh}
\end{figure*}

 The results of simulations performed using the setup described and a 2:1 density ratio are shown in Figure~\ref{fig:kh} for four different cases (top to bottom), with results shown at $\tau_{KH} = 1, 2, 4, 6, 8$ and $10$ (left to right). The four cases are, as indicated in the figure: 1) (top) using no artificial viscosity or thermal conductivity; 2) using artificial viscosity, applied uniformly with $\alpha = 1, \beta = 2$; 3) using the RT01 method with artificial viscosity applied using  the switch described in \S\ref{sec:switches}; 4) using the usual SPH formulation applying thermal conductivity with $\alpha_{u} = 1$ and our new signal velocity; and finally 5) as in case 4) but including both thermal conductivity and artificial viscosity (applied using switches). Case 2 thus represents the SPH formulation currently most widely used (that is, using artificial viscosity with $\alpha = 1$, $\beta = 2$) whilst case 5 represents the formulation we are proposing for general use.
 
  The differences between the five cases presented in Figure~\ref{fig:kh} are striking. Firstly, in the absence of any dissipation (case 1, top row of Figure~\ref{fig:kh}), whilst there is some growth in the velocity perturbation, mixing between the two layers is inhibited by what can only be described as an ``artificial surface tension'' in the dense fluid which results in blobs concentrating into well-defined features. This is particularly obvious in an animation of the simulation where blobs of dense fluid can be seen to separate and condense into increasingly spherical ``bubbles'' as time advances. Adding artificial viscosity (case 2) does not improve the situation, but rather makes things worse by suppressing the initial growth in the velocity perturbation for $\tau_{KH} < 6$ whilst showing similar surface tension effects at later times. This is in agreement with the results found by \citet{agertzetal} where reducing the artificial viscosity resulted in an increased tendency for layers mix but does not remove the fact that SPH, in its standard form, clearly has a problem with Kelvin-Helmholtz instabilities across density jumps.

 Use of the RT01 formulation (third row of Figure~\ref{fig:kh}, here shown with viscosity applied using the Morris \& Monaghan switch) improves things to some extent, showing some Kelvin-Helmholtz-instability like features, though at the expense of considerable particle noise at the interface (a close-up of which is shown in the top left panel of Figure~\ref{fig:khzoom}).
 
  The results in case 4 (adding artificial thermal conductivity via the method discussed in \S\ref{sec:switches}) are substantially different. In this case there is a clear development of the classical Kelvin-Helmholtz instability (although still on a slower overall timescale than expected from analytic theory), with growth in the $\lambda = 1/6$ mode visible at $\tau_{KH} \sim 2$ which is subsequently overtaken by a growth in the $\lambda = 1/2$ mode at $\tau_{KH} \sim 6$. The relative growth rate of the two modes is, however, in good agreement with theory (that is, a factor of $\sim$3 difference). Finally, case 5 demonstrated that the dramatic improvement in the results found using the artificial thermal conductivity are not significantly changed by the turning the artificial viscosity back on (which has been applied exactly as in the shock tube test presented in Figure~\ref{fig:sodfinal}), largely because the \citet{mm97} switch (\S\ref{sec:switches}) is very effective at turning off the artificial viscosity where there is no compression.

\begin{figure*}
\includegraphics[width=\textwidth]{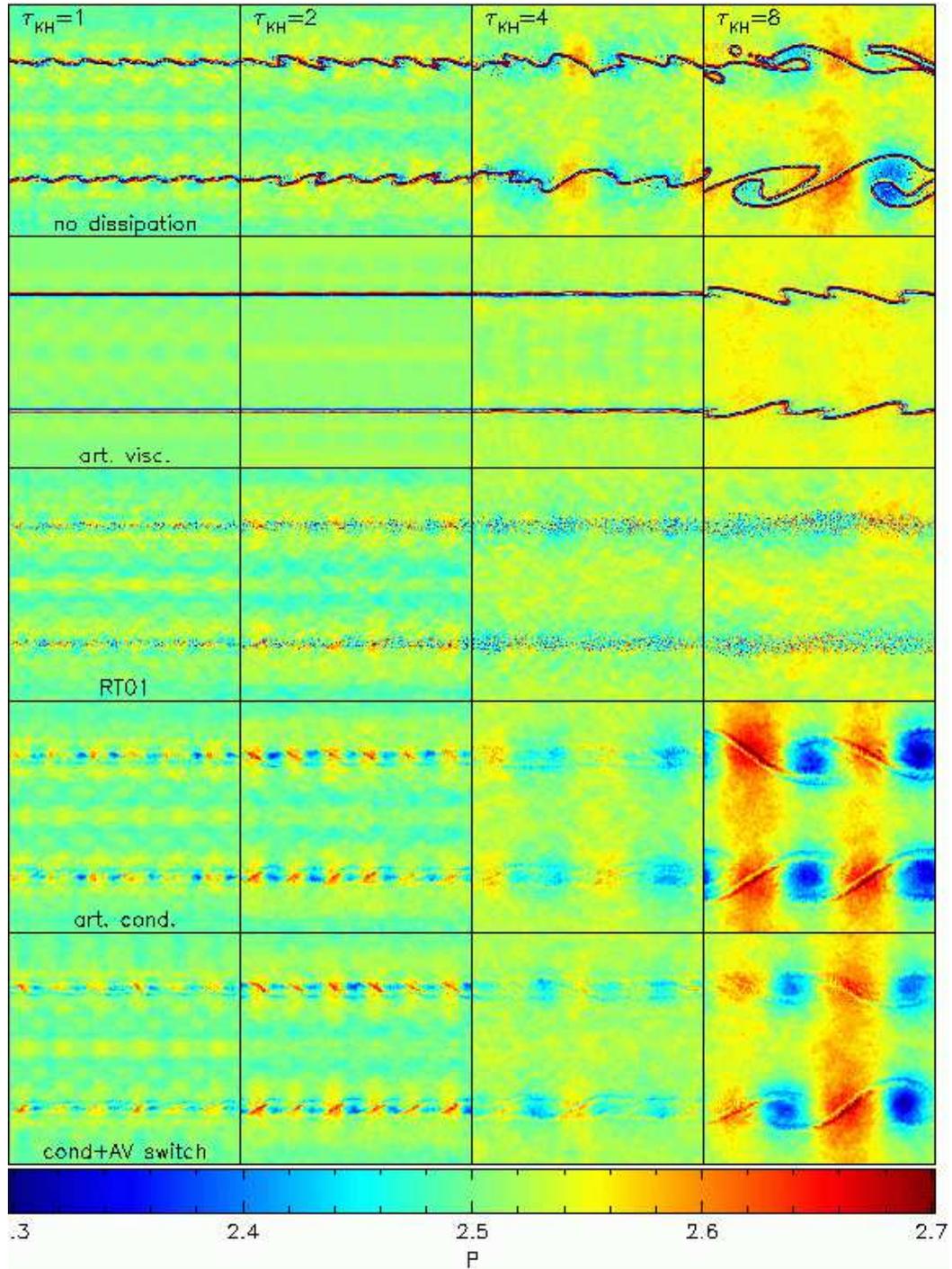}
\caption{Pressure distribution in the Kelvin-Helmholtz instability test shown in Figure~\ref{fig:kh}. Note in particular the regions of high and low pressure delineating the boundary between the two fluids in the absence of conductivity similar to the pressure blip observed around the contact discontinuity in Figure~\ref{fig:rhosum}.}
\label{fig:khpressure}
\end{figure*}

\begin{figure*}
\includegraphics[width=\textwidth]{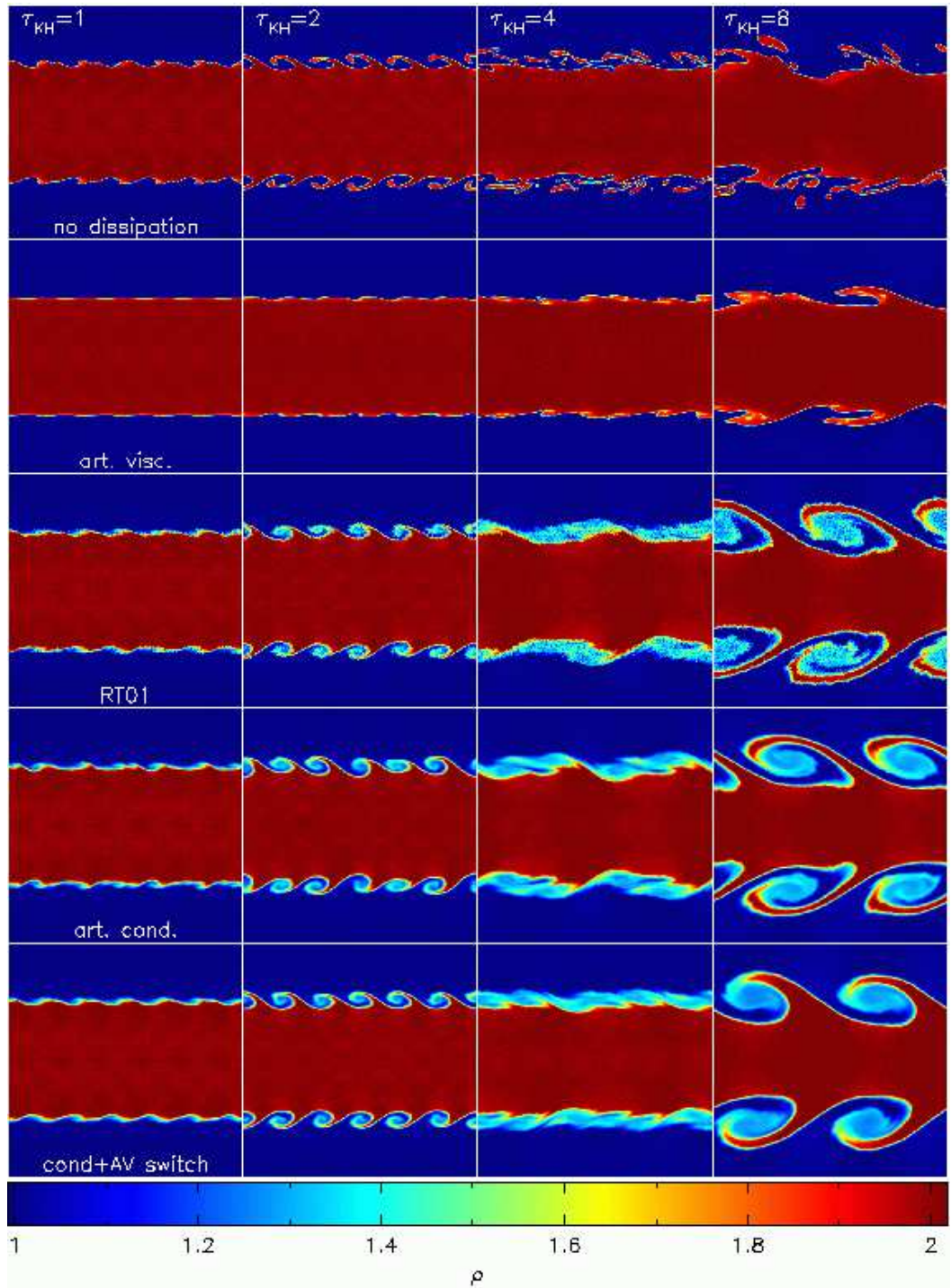}
\caption{Results of the Kelvin-Helmholtz instability test using a density ratio of 2:1 as in Figure~\ref{fig:kh} but here using an initial particle spacing of $\Delta = 1/512$ in the least-dense component. The results are similar to Figure~\ref{fig:kh}, namely that adding the artificial thermal conductivity term gives a dramatic improvement in SPH's ability to resolve the Kelvin-Helmholtz instability.}
\label{fig:kh_hires}
\end{figure*}

\begin{figure*}
\includegraphics[width=\textwidth]{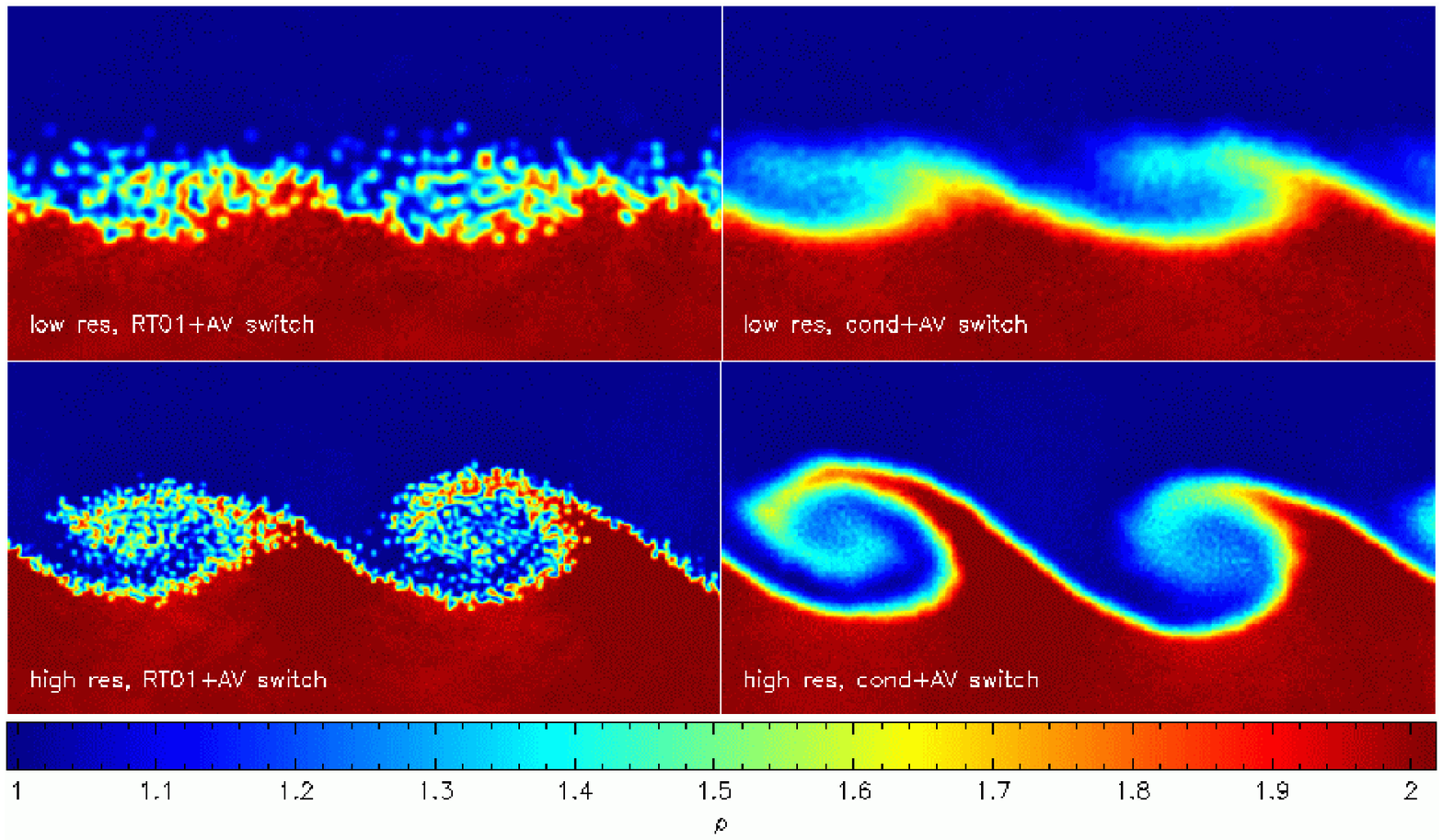}
\caption{Zoom up of selected panels from Figures~\ref{fig:kh} and \ref{fig:kh_hires} at $\tau_{KH} = 2$, highlighting the vortex rolls produced using the RT01 formulation (left panels) to a standard SPH formulation using our thermal conductivity term (right panels). The RT01 method works by effectively blurring the discontinuity, which though it helps to resolve the Kelvin-Helmholtz instability, also results in considerable particle noise at the interface.}
\label{fig:khzoom}
\end{figure*}

 An understanding of these results is provided by a plot of the pressure distribution in the same set of simulations, shown in Figure~\ref{fig:khpressure}. In cases 1 and 2 the boundary between the two fluids is clearly delineated by a ridge of high and low pressure. This pressure ``blip'' at the interface is exactly analogous to that observed in the one dimensional shock tube problems (e.g. Figure~\ref{fig:rhosum}) and is similarly cured by the same solution. Using the RT01 method (case 3) the pressure boundary is much less delineated, though considerably noisier, giving an improved mixing between the layers compared to the standard SPH formulation, at the expense of particle noise. Applying the usual SPH method with our artificial thermal conductivity term (cases 4 and 5) results in a smooth interface and correspondingly good mixing between the layers and a nice KH instability.

 To examine the effect of numerical resolution on the results, the five cases discussed above are presented at higher resolution (using $\Delta = 1/512$ in the least-dense component) in Figure~\ref{fig:kh_hires}. Comparison of case 1 (top row) shows that the artificial surface tension effect is not significantly modified by simply using more particles, although the size scale of the ``blobs'' and ``bubbles'' of dense fluid which break off into the less-dense component are somewhat smaller. Again, adding artificial viscosity acts to suppress the growth of velocity perturbations (case 2, second row), although in this case some perturbations are visible at earlier times compared to Figure~\ref{fig:kh}, suggesting that the effect of viscosity is lessened (which we expect since the artificial viscosity diffusion coefficient is linearly proportional to the particle spacing). The RT01 method also improves at this resolution, showing clear growth in both the $\lambda = 1/6$ mode and the $\lambda = 1/2$ mode (the latter not well resolved in the lower resolution RT01 run in Figure~\ref{fig:kh}), though the interface remains considerably noisy (see close-up shown in the lower left panel of Figure~\ref{fig:khzoom}).
 
  In case 4 (applying artificial thermal conductivity using our new $v_{sig}^{u}$) the Kelvin-Helmholtz instability is in this case well-resolved for both the $\lambda = 1/6$ and the $\lambda = 1/2$ modes, with well-resolved vortex rolls. The perturbation is also much stronger at earlier times (ie. $\tau_{KH} = 1$) compared to the lower resolution version in Figure~\ref{fig:kh}. Similar results are again found for case 5, indicating that our proposed ``generic formulation'' gives good results both for this test and on shock-tube problems. The results are also much cleaner than those obtained using the RT01 formulation, visible in Figure~\ref{fig:khzoom}), though their formulation has the merit of being non-dissipative.

\begin{figure*}
\includegraphics[width=\textwidth]{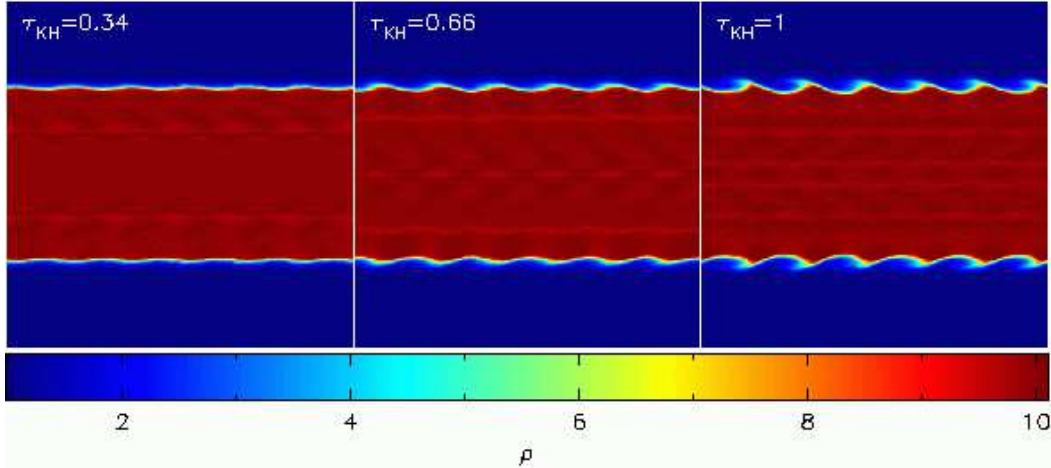}
\caption{Results of the Kelvin-Helmholtz instability test using a density ratio of 10:1 with an particle spacing of $\Delta = 1/256$ in the least-dense component, shown at the times corresponding to those shown in \citet{agertzetal} ($\tau_{KH} = 1/3, 2/3$ and $1$). The SPH calculations are in good agreement with the grid-based calculations presented in that paper, although the vortex rolls are slightly less well resolved due to the lower resolution employed in the low density fluid.}
\label{fig:kh10_agertz}
\end{figure*}
\begin{figure*}
\includegraphics[width=\textwidth]{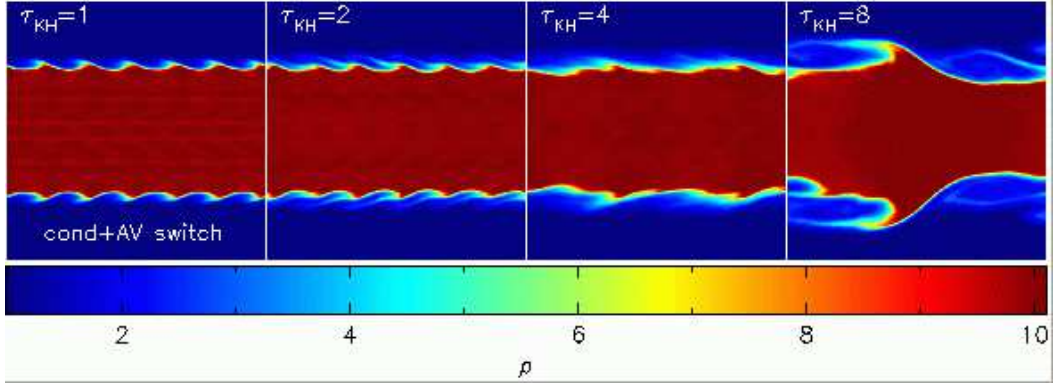}
\caption{Results of the Kelvin-Helmholtz instability test, as in Figure~\ref{fig:kh} but using a density ratio of 10:1 applying our new dissipation formulation.}
\label{fig:kh10}
\end{figure*}

   Whilst the tests presented above have been performed, for efficiency, using a density ratio of 2:1 between the two fluids, to ensure that our results are not dependent on the particular value of the density ratio employed, we have also performed a series of calculations using a density ratio of 10:1 (as in \citet{agertzetal}). The results of this test are shown at early times in Figure~\ref{fig:kh10_agertz} which may be directly compared to the corresponding figure in \citet{agertzetal}. In particular the results show very good agreement both with the analytic growth timescale and also with the grid-based calculations presented in \citet{agertzetal}, although the vortex rolls are slightly less well resolved in the present case due to the slightly lower resolution employed in the low density fluid (which differs by a factor of two from the two dimensional equivalent of the grid based calculations in the paper because we have used a box twice as large in $y$). The results at later times are shown in Figure~\ref{fig:kh10}. Again the development of the $\lambda=1/6$ mode is clearly seen followed by the $\lambda=1/2$ mode, exactly as in the 2:1 density ratio case (note that we have scaled the times in units of $\tau_{KH}$ in both cases). Thus we can confirm that our results represent a dramatic improvement over the SPH results discussed in \citet{agertzetal}.

\section{Discussion and Conclusions}
\label{sec:discussion}
 In this paper we have discussed several issues related to the treatment of discontinuities in SPH which have been shown to have a direct bearing on the recent problems highlighted by \citet{agertzetal} relating to simulating the Kelvin-Helmholtz instability across density jumps. The difference between integral and differential forms of the SPH equations was discussed, with the conclusion that whilst the continuity equation can be formulated in an integral form by using the density sum, the SPH momentum and energy equations are derived assuming that the equations are differentiable. This leads to the requirement for diffusion terms in both the momentum and energy evolution to capture discontinuities associated with each variable. Whilst discontinuities in momentum are treated by the application of artificial viscosity, discontinuities in thermal energy (e.g. at a contact discontinuity) were shown to require the application of an artificial thermal conductivity term, though \emph{this is largely ignored in many SPH calculations}. This leads to problems at contact discontinuities such as those found by \citet{agertzetal} because of the resultant discontinuous pressure profile.

  The reason for the ``artificial surface tension'' effect was observed to be a ``ridge'' or ``blip'' in the pressure at the interface between the two fluids in the standard SPH formulation (Figure~\ref{fig:khpressure}). This pressure ``blip'' is exactly the cause of the ``gap'' in the particle distribution discussed by \citet{agertzetal}. It is interesting therefore to note that in formulations of physical surface tension (e.g. \citealt{hls94}) the surface tension coefficient is directly related to the magnitude of the pressure jump at the interface by the Laplace-Young condition,
\begin{equation}
[P] = \tau \kappa,
\end{equation}
where $[P] = P - P'$ is the pressure jump normal to the interface, $\tau$ is the surface tension coefficient and $\kappa$ is the interfacial curvature.

 The presence of artificial surface tension in SPH can also be understood as a consequence of the exact advection of entropy by SPH particles (which as discussed in \S\ref{sec:sphlagrangian} is a differential rather than integral conservation law) using the following thought experiment (Springel 2005, private communication): Consider a two-phase distribution of SPH particles where each phase contains particles of different entropy placed in a close box in pressure equilibrium. Because entropy is locally conserved there is no difference between a configuration in which the two phases are separated into distinct spatial regions and a configuration in which the two are completely mixed. However if there is spurious energy associated with the interface (ie. the pressure is not continuous) then the unmixed configuration is energetically preferred, leading to a tendency for low entropy fluid to form discrete blobs exactly as observed in the Kelvin-Helmholtz tests presented in Figures~\ref{fig:kh} and \ref{fig:kh_hires}. The effect can be removed by enforcing pressure continuity across interfaces as we have proposed above, which in our case is enforced by adding a small amount of entropy mixing (ie. an artificial thermal conductivity) at the interface. Note that this argument applies equally to SPH formulations evolving the thermal or total energies since such formulations differ only in relation to timestepping provided the variable smoothing length terms have been accounted for, and even without these terms the errors in the entropy evolution are small.

 To remedy this problem we have presented new formulation for artificial thermal conductivity in SPH (\S\ref{sec:switches}) which acts to equalise pressure at a contact discontinuity but which minimises dissipation elsewhere. This term, when applied to the Kelvin-Helmholtz problem was found to give excellent results in good agreement with both analytic estimates and the grid-based calculations presented in \citet{agertzetal}. We expect this term to be very effective at turning off artificial thermal conductivity for hydrodynamic problems where pressure gradients always represent non-equilibrium features. However, for simulations where other physical effects are included (e.g. gravity) it is possible to have pressure gradients that are nonetheless in equilibrium because of additional forces in the system. Thus the signal velocity we have proposed may require some modification in this case (for example using the ``net pressure difference'' accounting for both pressure and gravity between the particle pair) although we do not envisage that it would be difficult to do so.

 It is important to note that problems with Kelvin-Helmholtz instabilities occur because of entropy gradients which are not treated correctly, \emph{not} density gradients as has often been assumed. Thus a corollary to our results is that no special treatment is required in order to capture Kelvin-Helmholtz instabilities in isothermal flows (as often used in astrophysics) since in this case no entropy gradient is present.

 It is worth briefly discussing alternative approaches to the problems described above which would be interesting to explore. The first is that, since the problem at contact discontinuities appears an ``artificial surface tension'' effect, a fruitful approach might be to try to formulate an ``inverse surface tension'' which counteracts the effect of the surface created by the gradient in particle number density in the manner of the SPH surface tension formulation recently introduced by \citet{huadams06}. The advantage would be that this would be a truly dissipation-less approach to tackling contact discontinuities, going some way towards actually accounting for the surface integral terms which have ``gone missing'' from the standard SPH formulation by the assumption of differentiability. A second approach would be to try to actually calculate the surface terms associated with discontinuities in momentum and thermal energy (for example \citet{kats01} presents a Lagrangian variational principle which incorporates discontinuities by retaining surface integral terms). Thirdly our expectation based on the results presented in \citet{inutsuka02} is that the formulation of Godunov-SPH which accounts for the surface integral terms should also give good results for the Kelvin-Helmholtz instability problem because of the continuity of pressure across the contact discontinuity.

 In summary, based on the fact that the Kelvin-Helmholtz instability is strongly inhibited across density jumps in ``standard'' SPH formulations (adopting only an artificial viscosity term) and the effectiveness with which our proposed signal velocity turns off artificial thermal conductivity where it is not needed (at least for non-self-gravitating calculations), we suggest that the ``generic formulation'' of dissipation terms presented in this paper should be generally adopted as standard procedure in SPH calculations.

\section*{Acknowledgments}
DJP is supported by a UK PPARC/STFC postdoctoral research fellowship. I would like to thank the many people who have encouraged me to write up the ideas presented in this paper, including Matthew Bate, Giuseppe Lodato, Jim Pringle, Thorsten Naab, Markus Wetzstein and Walter Dehnen. Extra thanks go to Joe Monaghan from whom I have learnt most of the ideas discussed in this paper and to Seung-Hoon Cha for useful discussions regarding Godunov-SPH. Thanks also to Mike Norman and Volker Springel for helpful discussions. Figures and exact solutions have been produced using SPLASH \citep{splashpaper}, a visualisation tool for SPH that is publicly available at http://www.astro.ex.ac.uk/people/dprice. 

\bibliography{sph}

\label{lastpage}
\end{document}